\documentclass[prd, preprint, longbibliography, 11pt]{revtex4-1}

\usepackage{amsmath}
\usepackage{mathtools}
\usepackage{amssymb}
\usepackage{setspace}
\usepackage{graphicx}
\usepackage{natbib}
\usepackage{float}
\usepackage{tensor}
\usepackage[utf8]{inputenc}
\usepackage{amsfonts}
\usepackage{braket}
\usepackage{esint}

\usepackage{breqn}
\usepackage{IEEEtrantools}

\usepackage{blindtext}
\usepackage{graphicx}
\usepackage{amsmath}
\usepackage{dirtytalk}

\begin{document}

\begin{center}
{ \Large \bf   Left-Right Symmetric Fermions and Sterile Neutrinos from Complex Split Biquaternions and Bioctonions}

\vskip 0.2 in

{\large{\bf Vatsalya Vaibhav$^{a1}$ and Tejinder P.  Singh$^{b2}$ }}

\medskip

{\it $^a$Indian Institute of Technology Kanpur, 208016, India}\\

{\it $^b$Tata Institute of Fundamental Research, Homi Bhabha Road, Mumbai 400005, India}\\

\bigskip

 \; {$^{1}$\tt vatsalya@iitk.ac.in},  \; {$^2$\tt tpsingh@tifr.res.in}

\bigskip



\end{center}

\centerline{\bf ABSTRACT}
\smallskip
\noindent In this article we investigate the application of complex split biquaternions and bioctonions to the standard model. We show that the Clifford algebras $Cl(3)$ and $Cl(7)$ can be used for making left-right symmetric fermions. Hence we incorporate right-handed neutrinos in the division algebras-based approach to the standard model. The right-handed neutrinos, or sterile neutrinos, are a potential dark-matter candidate. Using the division algebra approach, we discuss the left-right symmetric fermions and their phenomenology. We describe the gauge groups associated with the left-right symmetric model and prospects for unification through division algebras. We briefly discuss the possibility of obtaining  three generations of fermions and charge/mass ratios through the exceptional Jordan algebra $J_3(O)$ and the exceptional groups $F_4$ and $E_6$.

\bigskip

\section{Introduction}
\noindent The quaternions were initially introduced by Hamilton to explain rotations in three dimensions, and they form a non-commutative division algebra. In \cite{Gunaydin}, Gunaydin and Gursey proposed the use of octonions to understand quarks. Following them since, several authors have investigated the application of division algebras and Clifford algebras in the context of particle physics \cite{Gunaydin}-\cite{Wilson}. In the present paper, we build on the earlier work  obtaining particles from the left ideals of Clifford algebras \cite{Gunaydin, Furey1, Furey2, Furey3, Gillard, Stoica}. We use the Clifford algebras $Cl(3)$ and $Cl(7)$ to obtain a left-right symmetric model for fermions. The left-right symmetric model for fermions was introduced in 1975 by Senjanovic, Mohapatra, Pati, and Salam \cite{RN, GS, GS2, AM}; this model was not widely accepted back then because it predicted neutrinos to be massive. After the discovery of possible evidence for neutrino mass in 2002 \cite{SNO, Kamiokande}, the left-right symmetric model again gained attention. The L-R model accounts for neutrino masses using the see-saw mechanism \cite{GS2}, and proposes a right-handed neutrino with a significantly higher Majorana mass. The right-handed neutrino interacts through gravitation only (hence the name {sterile neutrino}) and hence is a potential candidate for dark matter \cite{Hall3}. The right-handed neutrino is expected to interact through the right-handed analog of weak force as proposed by the Pati-Salam model \cite{Pati}, but this force manifests itself only at high energies. 
The left-right symmetric model explains why charge-parity is conserved in strong interaction (the strong CP problem) and why the Higgs-coupling vanishes prior to the electroweak symmetry breaking \cite{Hall1, Hall2}. We discuss sterile neutrinos and the Higgs coupling in the context of division algebras. We also discuss  alternative approaches to the Pati-Salam model in the context of unification and propose the presence of gravity mediating bosons from division algebras in a pre-spacetime theory.

In their work \cite{Baylis, Baylis2, Baylis3}, Trayling and Baylis propose a geometrical approach for understanding the $Cl(7)$ algebra. They propose a higher dimensional Kaluza-Klein theory with four \say{external} spacetime dimensions accounting for rotations and boosts and four \say{internal} dimensions to explain the electroweak and color sector of the standard model. All these eight dimensions are naturally present in the octonionic chains or matrices in $\mathbb{R}[8]$ (see \cite{Furey1}) which make the $Cl(6)$ and $Cl(7)$ algebra. Trayling and Baylis also discuss right-handed sterile neutrinos and Higgs field coupling from $Cl(7)$. In this paper, we show that we can get two sets of fermions with opposite chirality from $Cl(7)$. This happens because $Cl(3)$ and $Cl(7)$ have two irreducible pinor groups; therefore we are able to break them as a direct sum of two copies of $Cl(2)$ and $Cl(6)$ respectively.

Wilson \cite{Wilson} observes the Pati-Salam gauge group as a $Spin(6)\times Spin(4)$ group coming naturally from the Clifford algebras $Cl(0,6)$ and $Cl(3,3)$ \cite{Wilson}. It is interesting to note that $Cl(7)$ which gives us two copies of $Cl(6)$ is giving us left-right symmetric fermions whereas the real algebras $Cl(0,6)$ and $Cl(3,3)$ are giving us the Pati-Salam gauge group. Boyle also discusses getting the Pati-Salam gauge group from the intersection of the two maximal subgroups of the exceptional group $E_6$ \cite{Boyle}. $E_6$ is the complexified version of $F_4$ which is the automorphism group of exceptional Jordan matrices $J_3(\mathbb{O})$. A relation between exceptional groups and the standard model symmetries has been investigated by authors in \cite{Baez2, Boyle, Todorov1, Todorov2, Todorov3, lisi, will, will2, tp3}.

In the next sub-sections, we briefly describe the mathematics underlying the physics of the present paper. For a more detailed discussion on mathematics please refer to \cite{Dixon, Furey1, Baez, I, Dray2, Penn}.

\subsection{Division Algebras}
A division algebra is an algebra in which every non-zero element has a multiplicative inverse. There are only four normed division algebras namely $\mathbb{R}, \mathbb{C}, \mathbb{H}$, and $\mathbb{O}$. $\mathbb{R}$ is the algebra of real numbers, $\mathbb{C}$ is the algebra of complex numbers, $\mathbb{H}$ is the algebra of quaternions, and $\mathbb{O}$ is the algebra of octonions. Quaternions are non-commutative in nature, whereas octonions are neither commutative nor associative. 

$\mathbb{H}$ is the algebra of the quaternions made up of one real unit and three imaginary units $i, j, k$. The quaternions have the following multiplication rule:
\begin{align}
    i^2 = j^2 = k^2 = ijk = -1\\
    ij + ji = jk + kj = ki + ik = 0\\
    ij = k, jk = i, ki = j 
\end{align}
The octonions make the non-commutative, non-associative division algebra $\mathbb{O}$. The octonions are made up of one real unit and 7 imaginary units $e_i^2 = -1$ for i = 1, 2,..., 7. Just like quaternions, the imaginary units of octonions anti-commute with each other. The multiplication of octonions is given by the following diagram known as the Fano plane.
\begin{figure}[h]
\centering
\includegraphics[width=7.5cm]{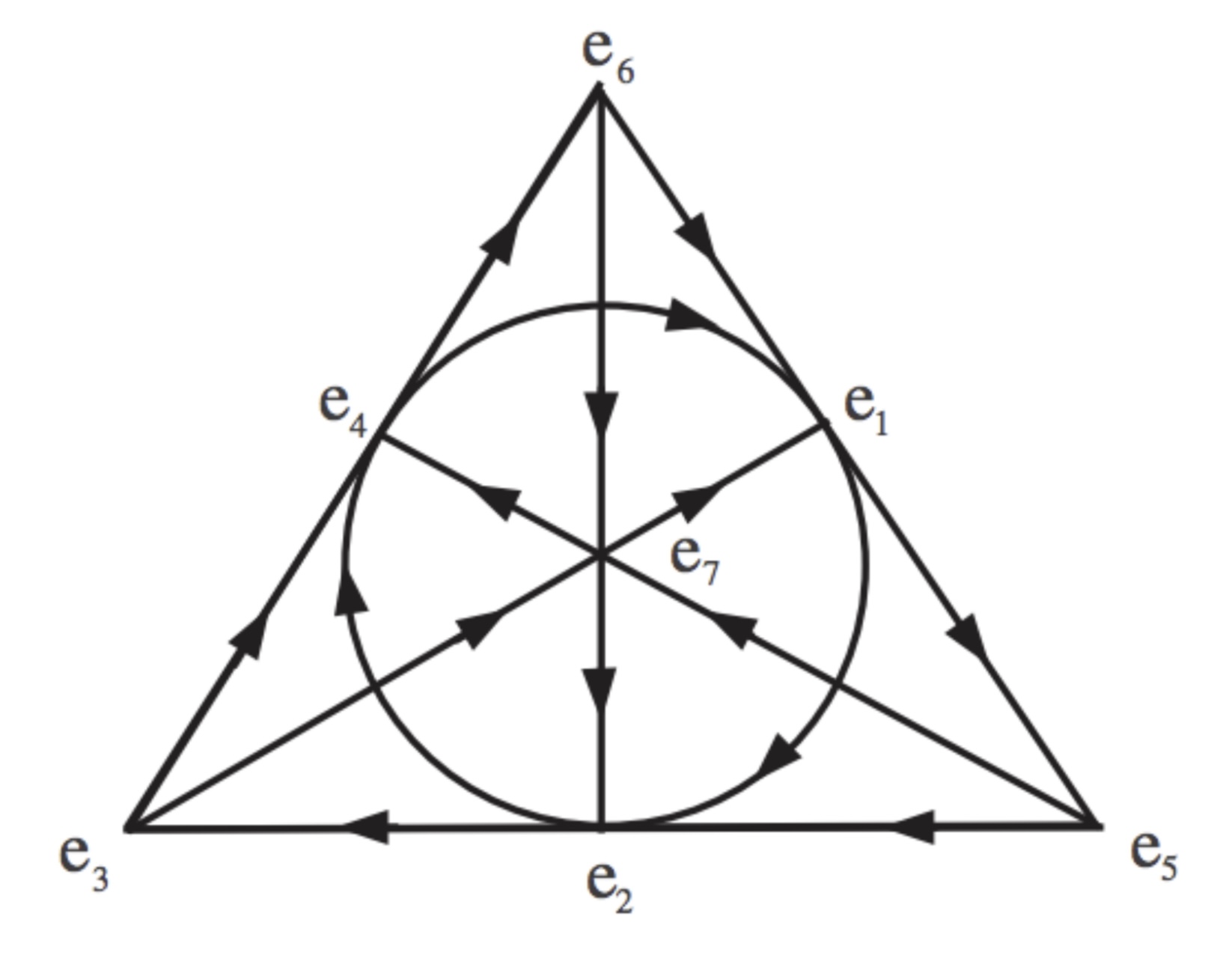}
\caption{The Fano plane}
\end{figure}
There are seven quaternionic subsets in the Fano plane, given by the three sides of the triangle, the three altitudes, and the incircle. Multiplication of points lying along a quaternionic subset in the cyclic order (as per the arrow) is given by $e_ie_j = e_k$, whereas $e_je_i = -e_k$.

It can be checked using the above multiplication laws that the octonions are not associative, but we can make an associative algebra by defining octonionic chains from octonions as is shown in \cite{Furey1}. For more details please refer to \cite{Dixon, Furey1, Baez}.

\subsection{Clifford Algebras}
In 1930 Jivet \cite{Jivet} and  Sauter \cite{Fritz1} showed that spinors are left ideals of matrix algebra. In 1947 the mathematician Marcel Riesz showed that spinors are minimal left ideals of the Clifford algebra \cite{Riesz}. Several authors have drawn a correspondence between the left ideals of Clifford algebras and the fermions from a standard procedure based on Witt decomposition proposed by Ablamowicz \cite{Ablamowicz, Furey1, Furey2, Furey3, Gillard, Stoica}. The Clifford algebras are so closely related to the division algebras that Clifford Algebras were introduced as an extension of the quaternions \cite{Clifford}.

A Clifford algebra over the field $\mathbb{R}$ is an associative algebra written as $Cl(p,q)$ such that there is a $n = p + q$ dimensional vector space $V = \{e_1, e_2, ..., e_n\} $ satisfying:
\begin{equation}
    \{e_i, e_j\} \equiv e_ie_j + e_je_i = 2\eta_{ij}\mathbf{I} 
\end{equation}
Here the vector space $V$ is called the generating space of $Cl(p,q)$. $\eta_{ij} = 0$ if $i \neq j$, $\eta_{ii} = 1$ for $i = 1,..., p$, $\eta_{ij} = -1$ for $i = p+1,...., p+q$. There will be an identity vector in the algebra, such that $a.1 = 1.a = a$, $\forall a \in Cl(p,q)$. The elements of a Clifford algebra can be constructed by multiplication of generating vectors, therefore, it can be checked that the dimension of the Clifford algebra $Cl(p,q)$ will be $\sum_{i=0}^{n} {}^nC_i = 2^n$.

We present some examples now; the $Cl(0,0)$ algebra will correspond to the real numbers $\mathbb{R}$. There are no generating vectors, just the trivial identity vector in this case. $Cl(0,1)$ will correspond to the complex numbers, here we have only one generating vector \say{$i$} with $i^2 = -1$. For $Cl(0,2)$ we need two generating vectors $e_1^2 = -1$ and $e_2^2 = -1$. The algebra will then be $Cl(0,2)$ = $\{1, e_1, e_2, e_1e_2\}$, we identify this algebra with the division algebra of quaternions $\mathbb{H}$. Therefore from the first three Clifford algebras, we obtain the division algebras of $\mathbb{R}, \mathbb{C}$, and $\mathbb{H}$. The octonions $\mathbb{O}$ do not make Clifford algebras naturally because of their non-associativity but they can be used to make the associative algebra $Cl(6)$ by introducing the octonionic chains, as is shown by Furey \cite{Furey1}.

The next Clifford algebra in this series is $Cl(0,3)$, the generating vectors will be $e_1^2 = -1, e_2^2 = -1$, and $e_3^2 = -1$. The Clifford algebra $Cl(0,3)$ will have eight elements $\{1, e_1, e_2, e_3, e_1e_2, e_2e_3, e_3e_1, e_1e_2e_3\}$. We can break this algebra into a sum of two quaternionic parts:
\begin{equation}
(1, e_1, e_2, e_1e_2), \quad(e_1e_2e_3, e_2e_3, e_3e_1, e_3)
\end{equation}
It is important to understand that $e_3 \neq e_1e_2$, unlike in the case of quaternions. We can identify the left set as the quaternions; in the right set $(e_1e_2e_3)^2 = 1$ whereas the squares of the other three elements are -1. If we denote $e_1e_2e_3$ as $\omega$ then the right set becomes $\omega(1, -e_1, -e_2, -e_1e_2)$. Thus the right set is $\omega$ times the quaternions. $\omega$ here is a split complex number; analogous to $i$ which squares to -1, the split complex number squares to 1 but is neither 1 nor -1. Thus, the algebra $Cl(0,3)$ is called the split-biquaternions, as termed by Clifford himself \cite{Clifford}. It can be written as $\mathbb{C}' \otimes \mathbb{H}$, where $\mathbb{C}' \equiv (1, \omega)$. Notice that $\mathbb{C}' \otimes \mathbb{H} \cong \mathbb{H} \oplus \mathbb{H}$, however, it is important to assert that the set involving $\omega$ is not precisely quaternionic.

Next, we move on to Clifford algebras on the complex field. The Clifford algebra $Cl(n)$ is defined on the complex field by an $n$ dimensional vector space $V = \{e_1,....,e_n\}$ such that:
\begin{align}
    \{e_i, e_j\} = 0,\quad i \neq j\\
    e_i^2 = 1.
\end{align}
It is interesting to note that the Clifford algebra $Cl(n)$ on the complex field can be obtained from the Clifford algebra $Cl(p,n-p)$ on the real field by the following relation: $Cl(n) = \mathbb{C} \otimes Cl(p,n-p)$ where $0 \leq p \leq n$.

Therefore, we note that we can get the algebra $Cl(3)$ from complexification of $Cl(0,3)$, and the final algebra would be complex split-biquaternions $\mathbb{C} \otimes (\mathbb{H} \oplus \omega\mathbb{H})$. The Clifford algebra $Cl(0,7)$ is the algebra of split $8 \times 8$ real matrices, $\mathbb{R}[8] \oplus \omega\mathbb{R}[8]$. On complexifying we will get $Cl(7)$ which is $\mathbb{C}[8] \oplus \omega\mathbb{C}[8]$. It has been shown by Furey in \cite{Furey1} that $Cl(6) = \mathbb{C}[8] \cong \overleftarrow{\mathbb{C}} \otimes \overleftarrow{\mathbb{O}}$. Here $\overleftarrow{\mathbb{C}} \otimes \overleftarrow{\mathbb{O}}$ is the algebra of complex octonionic chains, defined as a series of maps acting on a function $f \in \mathbb{C}\otimes\mathbb{O}$ from left to right. Therefore $Cl(7) \cong \overleftarrow{\mathbb{C}} \otimes \overleftarrow{\mathbb{O}} + \omega\overleftarrow{\mathbb{C}} \otimes \overleftarrow{\mathbb{O}}$, the complex split bioctonions.

It is interesting to note that such splitting of the algebra is present only in $Cl(0,3 + 4n)$. Both of these algebras have two irreducible representations. The irreducible representations of the complex Clifford algebra $Cl(n)$ are called pinors. It is worth noting that for $Cl(3)$ and $Cl(7)$ if one of the pinor group is left-handed, the other will be right-handed, this can be checked from the fact that the split $\omega\mathbb{H}$ is written as $\omega(1, -e_1, -e_2, -e_1e_2)$ and the signs of the imaginary coordinates have flipped. For a more extensive analysis of Clifford algebras please refer to \cite{Baez, Penn}.

In this section, we briefly discussed the mathematics required for this paper. We will subsequently see how we can use split complex biquaternions and bioctonions to obtain fermions in the left-right symmetric generalization of the standard model.

\section{Complex Quaternions $Cl(2)$}
The $Cl(2)$ algebra can be obtained by complexification of $Cl(0,2)$ and hence it is identified with complex quaternions $\mathbb{C} \otimes \mathbb{H}$. Complex quaternions can give us one generation of leptons as has been shown in \cite{Furey1} with the $U(1)_{em}$ symmetry coming from the number operator. In this section, we give a brief review of the results in \cite{Furey1} in the context of $Cl(2)$.

The generating vector space of $Cl(2)$ algebra is $W = \{ie_1, ie_2\}$. A subspace $U \subset W$ is called a maximal totally isotropic subspace (MTIS) of $W$, if:
\begin{equation}
    \{\alpha_i, \alpha_j\} = 0 \quad \forall \alpha_i \in U 
\end{equation}
We can see that the MTIS for $Cl(2)$ will be one dimensional with either the element $\alpha = \frac{e_1 + ie_2}{2}$ or $\frac{-e_1 + ie_2}{2}$. The primitive idempotent $V$ of this algebra can be defined as $\alpha \alpha^{\dagger}$. We can now create minimal left ideals of $Cl(2)$ by left multiplication of $\mathbb{C} \otimes \mathbb{H}$ on the idempotent. The obtained left ideals will be Weyl spinors of distinct chirality and spin.

If we use $\alpha = \frac{e_1 + ie_2}{2}$ to make the idempotent the obtained Weyl spinor will be:
\begin{equation} \label{1}
\psi_R = \epsilon^{\downarrow \uparrow}\alpha^{\dagger}V + \epsilon^{\uparrow \uparrow}V
\end{equation}
where $\epsilon^{\downarrow \uparrow}, \epsilon^{\uparrow \uparrow} \in \mathbb{C}$, $V$ is the idempotent. This right-handed Weyl spinor $\psi_R$ can be interpreted as a superposed state of $V$ and $\alpha^{\dagger}V$, where $V$ and $\alpha^{\dagger}V$ are two different leptons.
The difference between these two leptons is evident from the number operator 
\begin{equation}
    N = \alpha^{\dagger}\alpha
\end{equation}
The spinors $V$ and $\alpha^{\dagger}V$ are eigenvectors of this Hermitian number operator with eigenvalues 0 and 1 respectively. These eigenvalues correspond to the charge of the leptons. Therefore $V$ will be interpreted as the neutrino and $\alpha^{\dagger}V$ will be interpreted as the charged lepton.

We can similarly obtain a left handed Weyl spinor $\psi_L$ if we use the MTIS $\alpha = \frac{-e_1 + ie_2}{2}$. The idempotent obtained now will be $V^*$. Instead of $\alpha^{\dagger}$, we will now use $\alpha$ to obtain the excited state from the idempotent. The obtained left-handed Weyl spinor is
\begin{equation} \label{2}
\psi_L = \epsilon^{\uparrow \downarrow} \alpha V^* + \epsilon^{{\downarrow \downarrow}}V^*
\end{equation}
where $\epsilon^{\uparrow \downarrow}, \epsilon^{\downarrow \downarrow} \in \mathbb{C}$. Individually we can identify $V^*$ and $\alpha V^*$ as a neutrino and a charged lepton. The left-handed leptons will be the anti-particles of right-handed leptons and will be related through complex conjugation $*$.

The $SL(2,\mathbb{C})$ symmetry responsible for Lorentz invariance of vectors and spinors is already present in the $Cl(2)$ algebra. We can write a 4-vector in Minkowski spacetime using the quaternions as follows:
\begin{equation} \label{3}
    V = v_0 + v_1e_1 + v_2e_2 + v_3e_1e_2
\end{equation}
Any number $s$ in the $\mathbb{C} \otimes \mathbb{H}$ algebra which doesn't have a component along the real quaternionic direction can be shown to be the generator of the Lorentz algebra \cite{Furey1}. For $s = s_1e_1 + s_2ie_1 + s_3e_2 + s_4ie_2 + s_5e_1e_2 + s_6ie_1e_2$, the Lorentz operator $e^{is}$ will bring about boosts and rotations. Notice that $ie_1, ie_2, ie_1e_2$ are the Pauli matrices and the subgroup $SU(2)$ is already present in $SL(2,\mathbb{C})$. This SU(2) is the usual spin group responsible for rotational symmetry in 3-D space.

\section{Complex Split Biquaternions $Cl(3)$}
$Cl(3)$ algebra can be obtained by complexification of $Cl(0,3)$ algebra. We have seen in Section I that $Cl(0,3)$ is the split biquaternions, therefore we will call $Cl(3)$ as complex split biquaternions $\mathbb{C} \otimes \mathbb{C}' \otimes \mathbb{H}$. Recall that we wrote $Cl(0,3)$ as $\mathbb{C}' \otimes \mathbb{H} $, this is isomorphic to $\mathbb{H} \oplus \mathbb{H}$, and the $\omega$ is usually left out. However, we will subsequently see the physical importance of $\omega$. It seems that the symmetry laws will be the same for $\mathbb{H}$ and $\omega \mathbb{H}$ because the multiplication in any Lie algebra has two terms and the $\omega$ will get squared to one. Despite the fact that the two spaces will have similar symmetry laws, the two spaces are different in terms of chirality and $\omega$ can play a crucial role in understanding the constant interaction of left-right fermions with the Higgs.

In section I, we wrote the $Cl(0,3)$ algebra as a sum of two sets:
\begin{equation}
\quad (1, e_1, e_2, e_1e_2), \quad \omega(1, -e_1, -e_2, -e_1e_2)
\end{equation}
It is important to note that the two sets have opposite parity. If we create leptons from the two complex quaternions in $Cl(3)$ we will get two sets of leptons with opposite chirality.

The MTIS for the left set of complex quaternions will be either $\alpha = \frac{e_1 + ie_2}{2}$ or $\frac{-e_1 + ie_2}{2}$. As shown in \cite{Furey1} the particles and anti-particles created will be
\begin{align}
    \overline{\mathcal{V}}_{R} &= \frac{1 + ie_1e_2}{2} & \mathcal{V}_{L} &= \frac{1 - ie_1e_2}{2}\\
    e^{+}_{R} &= \frac{-e_1 + ie_2}{2} &
    e^{-}_{L} &= \frac{-e_1 - ie_2}{2} 
\end{align}
This way we have created the left-handed neutrino $\mathcal{V}_{L}$ and electron $e^{-}_{L}$ along with their right-handed anti-particles; the anti-neutrino $\overline{\mathcal{V}}_{R}$ and positron $e^{+}_{R}$. The particles and anti-particles are related to each other through complex conjugation $*$.

Similarly, we can get leptons from the right set of complex quaternions. The MTIS will be either $\alpha = \omega(\frac{-e_1 - ie_2}{2})$ or $\omega(\frac{e_1 - ie_2}{2})$. The particles and anti-particles created will be
\begin{align}
    \overline{\mathcal{V}}_{L} &= \frac{1 + ie_1e_2}{2} & \mathcal{V}_{R} &= \frac{1 - ie_1e_2}{2} 
   &  e^{+}_{L} &= \omega\frac{(e_1 - ie_2)}{2} &
    e^{-}_{R} &= \omega\frac{(e_1 + ie_2)}{2} 
\end{align}
As stated before the particles created from the right set of complex quaternions have opposite chirality to the particles created from the left set of complex quaternions. Therefore, we now have the right-handed neutrino $\mathcal{V}_{R}$ and the right-handed electron $e^{-}_{R}$, along with their left-handed anti-particles; the anti-neutrino $\overline{\mathcal{V}}_{L}$ and positron $e^{+}_{L}$. In section V we show why the left-handed fermions transform only under the $SU(2)_L$ gauge group and the right-handed fermions transform only under the $SU(2)_R$ gauge group. We note from Eqns. (14) and (16) that the left-handed neutrino and the right-handed neutrino are algebraically the same.

It has been pointed out in \cite{Furey1} that parity transformation can be brought by $e_i \rightarrow -e_i$, our result here is consistent with this fact.
The generator for $U(1)_{em}$
\begin{equation}
    Q = \alpha^{\dagger}\alpha
\end{equation}
is present in $Cl(3)$ and provides a charge to the leptons of both sectors.

The \say{physical} electron that we detect in our experiments is in constant interaction with the Higgs boson. All the leptons created from the left set of quaternions are either left-handed particles or their right-handed anti-particles, therefore they have a weak hypercharge and interact with the weak bosons through $SU(2)_L$ symmetry. The leptons created from the right set of complex quaternions on the other hand cannot interact with the weak bosons. The Higgs acts as a source and sink for hypercharge and changes the right-handed electron to left-handed, and the left-handed electron to right-handed.

It is worth noting that the $\omega$ in $Cl(3)$ maps the left-handed electron to the right-handed electron and conversely the right-handed electron to the left-handed electron. It is also worth noting that $\omega$ is self-adjoint
\begin{equation}
    (e_1e_2e_3)^{\dagger} = e_1e_2e_3
\end{equation}
Therefore, there is a global $U(1)$ symmetry associated with $\omega$, similar to the global $U(1)$ symmetry associated with the Higgs boson in the standard model. $\omega$ is an algebraic element of $Cl(0,3)$ that corresponds to the pseudoscalar in the Clifford algebra, therefore it might be used to represent the Higgs.

\section{Complex Split Bioctonions $Cl(7)$}
The relationship between quarks and division algebras has intrigued many authors before \cite{Gunaydin, Dixon, Furey1, Furey2, Furey3, Gillard, Baylis, Stoica}. Furey introduced the octonionic chain algebra to relate the non-associative octonions with $Cl(6)$. We have already defined the octonionic chains $\overleftarrow{\mathbb{C}} \otimes \overleftarrow{\mathbb{O}}$ in Section I. Throughout our analysis, we will take the function $f \in \mathbb{C}\otimes\mathbb{O}$ to be 1, so our octonionic chains act on 1.

The generators for $Cl(6)$ will be $\{i\overleftarrow{e_1}, i\overleftarrow{e_2}, i\overleftarrow{e_3}, i\overleftarrow{e_4}, i\overleftarrow{e_5}, i\overleftarrow{e_6}\}$. We are dropping the left arrow over octonions now, but throughout our analysis, we will be working with octonionic chains. The MTIS will be 3-dimensional with the following elements:
\begin{equation}
    \alpha_1 = \frac{-e_5 + ie_4}{2}, \quad \alpha_2 = \frac{-e_3 + ie_1}{2}, \quad \alpha_3 = \frac{-e_6 + ie_2}{2}
\end{equation}
If we define $\Omega = \alpha_1\alpha_2\alpha_3$, then the idempotent will be $\Omega\Omega^{\dagger} = \alpha_1\alpha_2\alpha_3\alpha_3^{\dagger}\alpha_2^{\dagger}\alpha_1^{\dagger}$. On left-multiplying the idempotent with elements of MTIS we obtain our excited states:
\begin{align}
    \overline{\mathcal{V}} = \Omega\Omega^{\dagger} = \frac{ie_7 + 1}{2}\\
    V_{ad1} = \alpha_{1}^{\dagger}\mathcal{V} = \frac{e_5 + ie_4}{2}\\
    V_{ad2} = \alpha_{2}^{\dagger}\mathcal{V} = \frac{e_3 + ie_1}{2}\\
    V_{ad3} = \alpha_{3}^{\dagger}\mathcal{V} = \frac{e_6 + ie_2}{2}\\
    V_{u1} = \alpha_{3}^{\dagger}\alpha_{2}^{\dagger}\mathcal{V} = \frac{e_4 + ie_5}{2}\\
    V_{u2} = \alpha_{1}^{\dagger}\alpha_{3}^{\dagger}\mathcal{V} = \frac{e_1 + ie_3}{2}\\
    V_{u3} = \alpha_{2}^{\dagger}\alpha_{1}^{\dagger}\mathcal{V} = \frac{e_2 + ie_6}{2}\\
    V_{e+} = \alpha_{3}^{\dagger}\alpha_{2}^{\dagger}\alpha_{1}^{\dagger}\mathcal{V} = -\frac{(i + e_7)}{2}
\end{align}
The following generator for $U(1)_{em}$ provides charge to the quarks and leptons
\begin{equation}
    Q = \frac{\alpha_1^{\dagger}\alpha_1 + \alpha_2^{\dagger}\alpha_2 + \alpha_3^{\dagger}\alpha_3}{3}
\end{equation}
Therefore we get one generation of quarks and leptons from $Cl(6)$; the anti-particles will be related through complex conjugation $*$. As can be seen, each quark comes in three colours. It has been shown by authors \cite{Furey1, Gillard, Furey2, Furey3} that the algebra $\overleftarrow{\mathbb{C}} \otimes \overleftarrow{\mathbb{O}}$ already has the symmetry group $SU(3)$ present in it. The $SU(3)$ generators in terms of octonions are given by:
\begin{align}
    \Lambda_1 &= -\alpha_2^{\dagger}\alpha_1 - \alpha_1^{\dagger}\alpha_2 & \Lambda_5 &= -i\alpha_1^{\dagger}\alpha_3 + i\alpha_3^{\dagger}\alpha_1 \\
    \Lambda_2 &= i\alpha_2^{\dagger}\alpha_1 - i\alpha_1^{\dagger}\alpha_2 & \Lambda_6 &= \alpha_3^{\dagger}\alpha_2 - \alpha_2^{\dagger}\alpha_3 \\
    \Lambda_3 &= \alpha_2^{\dagger}\alpha_2 - \alpha_1^{\dagger}\alpha_1 & \Lambda_7 &= i\alpha_3^{\dagger}\alpha_2 - i\alpha_2^{\dagger}\alpha_3 \\
    \Lambda_4 &= -\alpha_1^{\dagger}\alpha_3 - \alpha_3^{\dagger}\alpha_1 & \Lambda_8 &= -\frac{(\alpha_1^{\dagger}\alpha_1 + \alpha_2^{\dagger}\alpha_2 - 2\alpha_3^{\dagger}\alpha_3)}{\sqrt(3)}
\end{align}
It can be checked using the charge generator that $\Omega$ has a charge -1, whereas $\Omega^{\dagger}$ has a charge 1. The right multiplication of $\Omega$ on the particles (Eqns. 21-28) changes their isospin from up to down, whereas the right multiplication of $\Omega^{\dagger}$ on the anti-particles changes their isospin from down to up. Therefore, $\Omega$ mimics the $W^-$ boson and $\Omega^{\dagger}$ mimics the $W^+$ boson.

It is worth noting that using $Cl(6)$ we can create one generation of fermions with distinct chirality. Left-handed particles from $Cl(6)$ will have right-handed anti-particles. We will now investigate $Cl(7)$ to see that it can be naturally written as a sum of two $\overleftarrow{\mathbb{C}} \otimes \overleftarrow{\mathbb{O}}$ with opposite parity. Throughout our analysis octonions should be treated as octonionic chains acting on the function $f = 1$; it is important to note that unlike in $Cl(6)$ $\overleftarrow{e_7} \neq \overleftarrow{e_1e_2e_3e_4e_5e_6}$. For our convenience we replace $\overleftarrow{e_1e_2e_3e_4e_5e_6}$ by $\overleftarrow{e_8}$. $Cl(0,7)$ can be made from two sets of octonions:
\begin{equation}
    (1, e_1, e_2, e_3, e_4, e_5, e_6, e_8) \oplus \omega_{\mathbb{O}}(1, -e_1, -e_2, -e_3, -e_4, -e_5, -e_6, -e_8) 
\end{equation}
Here $\omega_{\mathbb{O}} = \overleftarrow{e_1e_2e_3e_4e_5e_6e_7}$. Henceforth, $\omega$ is to be interpreted as $\omega_{\mathbb{O}}$ unless explicitly stated as $\omega_{\mathbb{H}}$ for $Cl(3)$. The above line of reasoning is not direct to see, but we can understand it in the following way. $Cl(0,7)$ is split $8\times 8$ real matrices ${\bf R}[8]\oplus {\bf R}[8]$, therefore $Cl(7)$ will be ${\bf C}[8]\oplus {\bf C}[8]$, the Clifford algebra $Cl(6)$ is also ${\bf C}[8]$ and is isomorphic to complex octonionic chains. Therefore $Cl(7)$ will be isomorphic to complex split octonionic chains. Just like $e_1e_2e_3$ commutes with all elements in $Cl(3)$, $\overleftarrow{e_1e_2e_3e_4e_5e_6e_7}$ will commute with all the elements of $Cl(7)$, and it squares to 1.\\\\
We can now make particles from the left set and the right set in a similar way as is done in $Cl(6)$. The MTIS for the left set will be $\alpha_1 = \frac{-e_5 + ie_4}{2}, \alpha_2 = \frac{-e_3 + ie_1}{2}, \alpha_3 = \frac{-e_6 + ie_2}{2}$. The idempotent will be $\Omega_L\Omega_L^{\dagger} = \alpha_1\alpha_2\alpha_3\alpha_3^{\dagger}\alpha_2^{\dagger}\alpha_1^{\dagger}$. The left-handed neutrino family and their right-handed anti-particles are:
\begin{align}
    \overline{\mathcal{V}} &= \frac{ie_8 + 1}{2} & \mathcal{V} &= \frac{-ie_8 + 1}{2}\\
    V_{ad1} &= \frac{(e_5 + ie_4)}{2} & V_{d1} &= \frac{(e_5 - ie_4)}{2}\\
    V_{ad2} &= \frac{(e_3 + ie_1)}{2} & V_{d2} &= \frac{(e_3 - ie_1)}{2}\\
    V_{ad3} &= \frac{(e_6 + ie_2)}{2} & V_{d3} &= \frac{(e_6 - ie_2)}{2}\\
    V_{u1} &= \frac{(e_4 + ie_5)}{2} & V_{au1} &= \frac{(e_4 - ie_5)}{2}\\
    V_{u2} &= \frac{(e_1 + ie_3)}{2} & V_{au2} &= \frac{(e_1 - ie_3)}{2}\\
    V_{u3} &= \frac{(e_2 + ie_6)}{2} & V_{au3} &= \frac{(e_2 - ie_6)}{2}\\
    V_{e+} &= -\frac{(i + e_8)}{2} & V_{e-} &= -\frac{(-i + e_8)}{2}
\end{align}
The MTIS for the right set will be $\alpha_1 = -\omega\frac{-e_5 + ie_4}{2}$, $\alpha_2 = -\omega\frac{-e_3 + ie_1}{2}$, $\alpha_3 = -\omega\frac{-e_6 + ie_2}{2}$. The idempotent will be $\Omega_R\Omega_R^{\dagger} = \alpha_1\alpha_2\alpha_3\alpha_3^{\dagger}\alpha_2^{\dagger}\alpha_1^{\dagger}$. The right-handed neutrino family and their left-handed anti-particles are:
\begin{align}
    \overline{\mathcal{V}} &= \frac{ie_8 + 1}{2} & \mathcal{V} &= \frac{-ie_8 + 1}{2}\\
    V_{ad1} &= \omega\frac{(-e_5 - ie_4)}{2} & V_{d1} &= \omega\frac{(-e_5 + ie_4)}{2}\\
    V_{ad2} &= \omega\frac{(-e_3 - ie_1)}{2} & V_{d2} &= \omega\frac{(-e_3 + ie_1)}{2}\\
    V_{ad3} &= \omega\frac{(-e_6 - ie_2)}{2} & V_{d3} &= \omega\frac{(-e_6 + ie_2)}{2}\\
    V_{u1} &= \frac{(e_4 + ie_5)}{2} & V_{au1} &= \frac{(e_4 - ie_5)}{2}\\
    V_{u2} &= \frac{(e_1 + ie_3)}{2} & V_{au2} &= \frac{(e_1 - ie_3)}{2}\\
    V_{u3} &= \frac{(e_2 + ie_6)}{2} & V_{au3} &= \frac{(e_2 - ie_6)}{2}\\
    V_{e+} &= \omega\frac{(i + e_8)}{2} & V_{e-} &= \omega\frac{(-i + e_8)}{2}
\end{align}
Therefore from the two sets on $Cl(7)$ we have one generation of left-right symmetric fermions. Both the right sector and the left sector have $SU(3) \times U(1)$ symmetry. The $SU(3)\times U(1)$ groups are identical for both the left and the right sector, more about this can be found in \cite{priyankk}.

Just as mentioned before, $\Omega_L$ and $\Omega_L^{\dagger}$ will play the role of $W^-$ and $W^+$ bosons respectively. On the contrary, $\Omega_R$ and $\Omega_R^{\dagger}$ will play the role of right-handed analogues of weak bosons. In the next section, we see why $SU(2)_L$ acts only on left-handed particles and their right-handed anti-particles whereas $SU(2)_R$ acts only on right-handed particles and their left-handed anti-particles.

It is worth noting that $\omega_{\mathbb{O}}$ in $Cl(7)$ can explain the interaction of left-right symmetric fermions with the Higgs similar to the $\omega_{\mathbb{H}}$ in $Cl(3)$ but with an addition of quarks. In case of up quarks however the $\omega$ gets squared to 1. The absence of $\omega$ for neutrinos and up quarks, and the fact that the ratio of the square root of the mass of the electron, up-quark, and down quark is $\frac{1}{3}: \frac{2}{3}: 1$ motivated the authors to talk about mass ratios for one generation of lepto-quarks prior to symmetry breaking, in Section VI.

\section{Why $SU(2)_L$ and $SU(2)_R$ are selective?}
Our universe seems to show some bias towards left-handed fermions in the sense that weak bosons interact only with left-handed fermions. In quantum field theory, this is accepted in an ad-hoc manner on experimental grounds. Furey in her work \cite{Furey1, Furey2, Furey3} showed why only left-handed fermions interact through $SU(2)_L$. In the Pati-Salam model, we have both $SU(2)_L$ and $SU(2)_R$, in this section we extend Furey's work to show why only right-handed fermions interact through $SU(2)_R$ and left-handed fermions interact through $SU(2)_L$.

The right action of $\mathbb{C}\otimes\mathbb{H}$ on $\mathbb{C}\otimes\mathbb{H}$ will look like
\begin{equation}
    (1, e_1, e_2, e_1e_2)\otimes_{\mathbb{C}}(1, e_1, e_2, e_1e_2)
\end{equation}
This will give us the $Cl(4)$ algebra, which we can generate using the vectors
\begin{equation}
    \{\tau_1ie_2, \tau_2ie_2, \tau_3ie_2, ie_1\}
\end{equation}
Here $\tau_1 = \Omega_L + \Omega_L^{\dagger}$, $\tau_2 = i\Omega_L - i\Omega_L^{\dagger}$, $\tau_3 = \Omega_L\Omega_L^{\dagger} - \Omega_L^{\dagger}\Omega_L$. Here $\Omega_L$ is borrowed from our previous section. We note that the MTIS in this case will be two dimensional spanned by
\begin{equation}
    \beta_1 = \frac{-e_1 + ie_2\tau_3}{2}, \quad \beta_2 = \Omega_L^{\dagger}ie_2
\end{equation}
The idempotent will be $\beta_1^{\dagger}\beta_2^{\dagger}\beta_2\beta_1 = \mathcal{V}_R$. We can obtain the right ideals by the right multiplication of $Cl(4)$ on this ideal. We identify the following particles
\begin{align}
    \beta_1^{\dagger}\beta_2^{\dagger}\beta_2\beta_1 = \mathcal{V}_R; \quad e^{-}_R = \mathcal{V}_R\beta_1^{\dagger}\beta_2^{\dagger}\\
    e^{-}_L = \mathcal{V}_R\beta_2^{\dagger}; \quad \mathcal{V}_L = \mathcal{V}_R\beta_1^{\dagger}
\end{align}
The $SU(2)$ symmetry can be generated by the following three generators
\begin{equation}
    T_1 = \tau_1\frac{1 + ie_1e_2}{2}, T_2 = \tau_2\frac{1 + ie_1e_2}{2}, T_3 = \tau_3\frac{1 + ie_1e_2}{2}
\end{equation}
It is interesting to note that the $SU(2)$ operators will annihilate the particles $\mathcal{V}_R$ and $e_R^-$ (eq. 54), whereas it will interchange
$e_L^-$ and $\mathcal{V}_L$ as isospin states. Therefore, $SU(2)_L$ acts on left-handed particles or conversely on right-handed anti-particles.

Similarly we can understand the right action of $\mathbb{C}\otimes\mathbb{H}$ on $\omega \mathbb{C}\otimes\mathbb{H}$. This can be written as
\begin{equation}
    (1, -e_1, -e_2, -e_1e_2)\otimes_C(1, e_1, e_2, e_1e_2)
\end{equation}
The algebra will again be $Cl(4)$, and the particles generated through similar analysis will be
\begin{align}
    \beta_1^{\dagger}\beta_2^{\dagger}\beta_2\beta_1 = \mathcal{V}_L; \quad e^{-}_L = \mathcal{V}_L\beta_1^{\dagger}\beta_2^{\dagger}\\
    e^{-}_R = \mathcal{V}_L\beta_2^{\dagger}; \quad \mathcal{V}_R = \mathcal{V}_L\beta_1^{\dagger}
\end{align}
The $SU(2)$ generators will now annihilate $\mathcal{V}_L$ and $e_L^-$ (eq. 54), whereas it will interchange
$e_R^-$ and $\mathcal{V}_R$ as isospin states. Therefore, $SU(2)_R$ acts on right-handed particles or conversely on left-handed anti-particles.

It is interesting to note that the right action of $\mathbb{C}\otimes\mathbb{H}$ on $\mathbb{C}\otimes\mathbb{H}$ gives us $Cl(4)$ whereas the right action of $\mathbb{C}\otimes\mathbb{H}$ on $Cl(3)$ will give us $Cl(5)$. $Cl(4)$ can be generated by complexification of the Dirac algebra ($Cl(1,3)$), which is generated by the Dirac matrices $\gamma_0, \gamma_1, \gamma_2, \gamma_3$. The Coxeter element of $Cl(5)$ can be written as $\gamma_5 = i\gamma_0\gamma_1\gamma_2\gamma_3$. $\gamma_5$ squares to 1, and anti-commutes with the other $\gamma$ matrices, it also acts as the parity operator for the spinors. We stress that while working with $Cl(3)$ we did not have to explicitly define a parity reversal operator, and the right action of $\mathbb{C}\otimes \mathbb{H}$ on $Cl(3)$ is directly giving us $Cl(5)$.

It is interesting to note that $Cl(5) \cong Cl(6)^+$, therefore we can extend this $Cl(5)$ algebra to $Cl(6)$. It is this $Cl(6)$ that Stoica talks about in his paper with respect to weak symmetry, \cite{Stoica}. Stoica, therefore, talks about two $Cl(6)$, one of the $Cl(6)$ has the group $SO(4)$, whereas the other $Cl(6)$ is the one Furey uses to make one generation of fermions with the symmetry group $SU(3)\times U(1)$. We discuss this in more detail in the next section, where we do a proper analysis of the automorphism group of octonions: $G_2$ and its subgroups \cite{tp1}.

\section{Implications: Three generations, Mass-ratios, and Gravity}

In this section, we discuss further work that can be carried out in this direction. We provide a brief sketch of our work on mass ratios of fermions and using trace dynamics on a pre-spacetime pre-quantum octonionic manifold. More detail about these ideas can be found here \cite{tp1, tp3, vvs}.

 The see-saw mechanism predicts that the right-handed sterile neutrino should be significantly heavier than the left-handed neutrino. Experiments such as  the neutrinoless double beta decay have been proposed to infer the possible existence of right-handed sterile neutrinos \cite{AM, GS2}. It is interesting to note that the neutrino is required to be a Majorana neutrino for explaining the see-saw mechanism. Our work on mass ratios of fermions also requires the neutrino to be Majorana \cite{tp3, vvs}.

The automorphism group of the exceptional Jordan matrices is $F_4$. The group has been discussed in terms of physics earlier, by authors \cite{Baez2, Boyle, Todorov1, Todorov2, Todorov3, tp1}. It is interesting to note that $F_4$ has two stabilizer groups, $H_1$ and $H_2$. $H_1$ is the stabilizer group of $J_3(\mathbb{C})$ in $J_3(\mathbb{O})$ and is given by $[SU(3)\times SU(3)]/\mathbb{Z}_3$. The other maximal subgroup $H_2$ is $Spin(9)$ and stabilizes the idempotents of $J_3(\mathbb{O})$. The intersection of these two maximal subgroups is $SU(3)\times SU(2)\times U(1)$ which is the gauge group of the standard model. For more details please refer \cite{Boyle, Todorov3, Baez2}. On complexifying the $F_4$ group we can get the $E_6$ group with complex representations. It has been pointed in \cite{Boyle} that $E_6$ has two maximal subgroups $\tilde{H_1} = [SU(3)\times SU(3)\times SU(3)]/\mathbb{Z}_3$, $\tilde{H_2} = Spin(10)$ and their intersection gives us $SU(3)\times SU(2)_R\times SU(2)_L\times U(1)$ which is the gauge-group for left-right symmetric model.

In \cite{vvs} we show that the eigenvalues of the exceptional Jordan matrices with fermionic entries give us the mass ratios for three generations of the standard model fermions. We propose obtaining the three generations of fermions through rotations of the fermionic representations of any one generation, this corresponds to the transformation of $J_3(\mathbb{O})$. Gillard and Gresnigt \cite{Gillard} have also worked on the three-generation problem using the division algebra-based approach along with Dray and Manogue \cite{Dray1}. The mass ratios obtained from the eigenvalues of $J_3(\mathbb{O})$ are summarized in Figure 2. It is interesting to note that the ratio of the square root of the mass of the electron, up quark, and down quark is 1:2:3. Could it be possible that prior to electroweak symmetry breaking the left-handed and right-handed particles existed in a combined state of leptoquarks with electric charge and square root mass charge? More details about this work can be found in \cite{vvs}.

The description of chiral fermions using complex split biquaternions and complex split bioctonions has led to significant further developments. We have proposed \cite{priyankk} that the $U(1)$ number operator associated with the right chiral fermions be identified with the square root of mass, $\pm\sqrt{m}$, which like electric charge, has both signs. One sign of square root of mass is assigned to matter particles (electron, down quark, up quark), and the other sign of square root mass to their anti-particles. Just like electric charge, the square root of mass is shown to be quantized in the ratio $(1/3, 2/3, 1)$ with these ratios respectively assigned to the electron, the up quark, and the down quark.  That is, with respect to the electric charge assignment from the $U(1)_{em}$ of the first $Cl(6)$, the position of the electron and the down quark have been flipped in the $U(1)_{grav}$ of the second $Cl(6)$, whereas the position of the up quark remains unchanged. This assignment of square-root mass has led to a derivation of the mass ratios of the three fermion generations, and a proposal for unification  of the standard model with pre-gravitation, based on an $E_8 \times E_8$ symmetry \cite{priyankk, sherry}.

\begin{figure}[h]
\centering
\includegraphics[width=10cm]{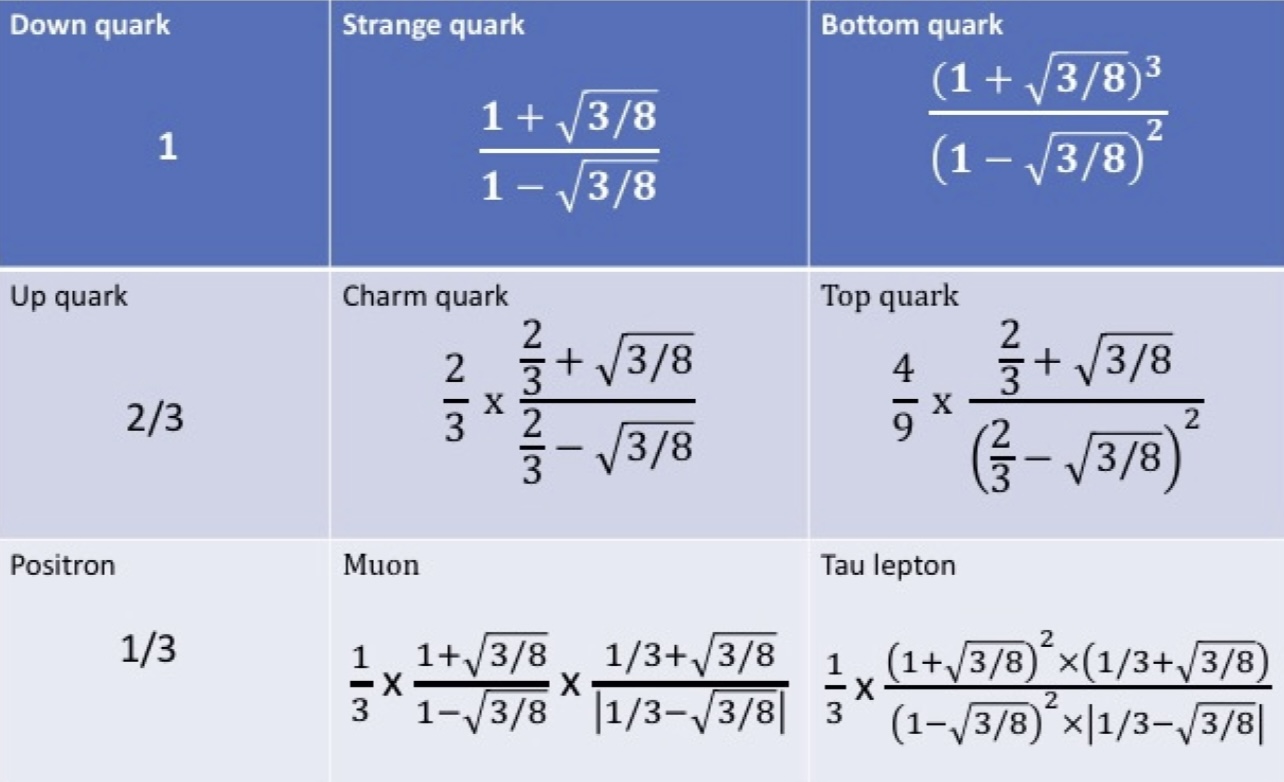}
\caption{Mass ratios of three generations of fermions}
\end{figure}

It is fascinating that we can talk about one generation of fermions and their electric charge from $G_2$, three generations of standard model fermions and their electric charge and square-root mass from $F_4$, $E_6$ being complexified $F_4$ gives us the left-right symmetric extension of the standard model.

There is a difference between the left-right symmetric model and the model that we propose in this paper through division algebras. The usual left-right symmetric model is a grand unified theory (GUT) on a Minkowski space-time background. We insist that unlike in quantum field theory, the division algebra approach to the standard model does not require a background spacetime. In \cite{tp1, tp2, tp3, Adler}, we advocate a pre-spacetime theory with a manifold mapped by octonions from which spacetime should emerge, we use trace dynamics (a collapse theory) to write the Lagrangian of the gauge fields (including gravity as one). Since we are working with trace dynamics, we do not impose any quantum conditions in an ad-hoc manner. In the division algebra-based approach the quantization of charge comes naturally from the number operator. Therefore, physics through division algebras can be a gateway to pre-spacetime, pre-quantum theories with the unification of all gauge fields.

It is important to stress that in a pre-spacetime theory, there is no spacetime, and the gauge fields lie in an abstract mathematical octonionic space, therefore there is no distinction between internal symmetries and spacetime symmetries. Since space-time and internal gauge symmetries arise as a result of spontaneous symmetry breaking of a unified symmetry, the Coleman-Mandula theorem prohibiting the mixing of internal symmetries with spacetime symmetries does not apply. (Also, we work with trace dynamics and not the $S$ matrix formalism). The result stated in \cite{Boyle, Todorov3, Baez2} has a strong interpretation. The automorphisms in maximal subgroup $H_2$ of $F_4$ preserves rotation of 10 dimensional spinor ($J_2(\mathbb{O})$) which form a subspace of $J_3(\mathbb{O})$, the intersection of the maximal subgroup $H_1$ and $H_2$ preserves spinors in 4-dimensional spacetime ($J_2(\mathbb{C})$). Therefore if we preserve 10-D spacetime in $J_3(\mathbb{O})$ and 4-D spacetime in the 10-D spacetime, the symmetry group we get is the standard model gauge group \cite{Baez2}.

Instead of claiming the right analogs of weak bosons from $SU(2)_R$, in \cite{priyankk, sherry}  we are using the $SU(2)_R$ for the chiral formulation of gravity. This approach has been used by other authors as well \cite{Woit}. In \cite{priyankk, sherry}, we propose three spin 1 Lorentz bosons instead of a spin 2 graviton. These Lorentz bosons are on an equal footing with the other  gauge bosons, and these bosons manifest themselves in a pre-spacetime, a pre-quantum theory called generalized trace dynamics \cite{tp1}-\cite{tp6}. 

Prior to symmetry breaking, we assume space-time to be 10 dimensional labeled by the octonions. Spacetime symmetry is associated with the action of $Sl(2, \mathbb{O})$ on $J_2[\mathbb{O}]$ the determinant of which gives a (1,9) signature. There is a unification of the Lorentz-weak symmetry with the electro-color symmetry. We have an L-R symmetric pre-quantum pre-spacetime theory in which the internal symmetries and the Lorentz symmetry of 4D spacetime have been unified via an 8D octonionic Kaluza-Klein theory. The dynamics is generalized trace dynamics, from which quantum field theory is emergent after symmetry breaking. In the unified theory, the concept of electric charge and square-root mass merge into one: charge-root-mass, which is quantized in units of 0, 1/3, 2/3, and 1, and comes with both signs, the negative sign being for anti-particles.

Lepto-quarks are bosons and will be described by the Clifford algebra $Cl(7)$ related to two copies of $Cl(6)$ and the split bioctonions. This underscores the importance of the split bioctonions studied in the present paper. Whereas in earlier sections the two copies of $Cl(6)$ both deal with electric charge as the quantum number, in the pre-theory one copy deals with electric charge, and the other with square-root mass, thus bringing gravitation within the framework of division algebras and Clifford algebras. 

\section{Conclusions}
We would like to conclude that the algebras $Cl(3)$ and $Cl(7)$ can be used to probe the left-right symmetric generalization of the standard model. The left-right symmetric model explains  the strong CP problem, the vanishing of Higgs coupling prior to symmetry breaking, the mass of the neutrino through a see-saw mechanism, and predicts a right-handed sterile neutrino which is a dark matter candidate. Further work needs to be done in this direction and a thorough geometric analysis of the octonionic space-time could shed further light on the work done in this paper.

\bigskip 

\noindent{\bf Acknowledgements:} It is a pleasure to thank Vivan Bhatt, Priyank Kaushik, Rajrupa Mondal,  and Robert Wilson for several helpful discussions. The authors are very thankful to the referees for their feedback and for playing a crucial role in improving this manuscript.

\end{document}